\documentclass[12pt]{article}
\usepackage{a4}

\begin{document}

\begin{titlepage}
\vfill

\vspace*{2cm}

~~
\vspace*{2cm}

~~

\begin{center}
{\Large\bf A Comment on a Recent Derivation of the Born Rule by Zurek}

\vspace*{1cm}

{\bf Cecilia Jarlskog}\\[0.3cm]
{Division of Mathematical Physics, Physics Department\\
LTH, Lund University, Lund, Sweden}
\end{center}
\vspace*{1cm}
\begin{abstract}
\noindent
The derivation of the Born rule by Zurek uses a ``splitting procedure'' where
a physical state is subdivided into a number of states. It is argued that in
quantum field theory, which encompasses quantum mechanics, such a procedure
would in general modify the physics.
\end{abstract}
\vfill

\end{titlepage}

\noindent 
Understanding the origin of Born rule \cite{born1926} in quantum mechanics is one of the greatest
challenges in physics. This rule is also valid in quantum field theory which encompasses quantum mechanics.
Briefly, suppose that we have a superposition of the form

\begin{equation}
 \vert \Psi > = \sum_{i=1}^{n} c_{i} \vert \psi_{i}>,
\end{equation}
where the $c_{i}$'s are constants and the 
state vectors $\vert \psi_{i}>$ form a complete orthonormal system. Moreover,
$\sum_{i=1}^{n} \vert c_{i}\vert^{2}=1$, to ensure that $\Psi$ is properly normalized.
Simply stated, the Born rule tells us that $\vert c_{i}\vert^{2}$ gives the probability
of finding $ \vert \Psi >$ in the state $\vert \psi_{i}>$. By now, 85 years after its inception,
no violation of this rule has been found.

\vspace*{.4cm}
\noindent
During the past several years Zurek has made great efforts to derive Born rule
by utilizing other postulates of
quantum mechanics. According to his own account, his work has been inspired by progress made 
in the domain of decoherence in quantum mechanics (see Ref. \cite{maxS} for an excellent 
pedagogical introduction to this subject). In a recent paper
\cite{zurek2011} Zurek gives an update of his derivation. He begins by 
considering a state of the form (which in this note we denote by $\vert F>$) 

\begin{equation}
\vert F> = \vert s_{1}> \vert e_{1}> + \vert s_{2}> + \vert e_{2}>,
\end{equation}
where $\vert s_{j}>$ and $\vert e_{j}>$, j=1,2, may be thought of as the two states
of a system S and its ``environment'' E.
We note in passing that $\vert F>$ is very special as it can be rewritten in the form
\begin{equation}
\vert F> = (\vert s_{1}>, ~\vert s_{2}>) \left( \begin{array}{cc}
1 & 0 \\ 0 &1 \end{array} \right) \left( \begin{array}{c} \vert e_{1}> \\ \vert e_2> \end{array} \right).
\end{equation}
$\vert F>$ may be looked upon as the scalar product of the two 
vectors (dropping for the moment the ket symbols) $\overline{s} =(s_{1},~ s_{2})^T$ and 
$\overline{e} =(e_{1}, ~e_{2})^T$. Thus
$F (= \overline{s}^T \overline{e}$) is a highly symmetric state. Indeed, it is 
invariant under rotations $R(\theta)$

\[ \overline{s} \rightarrow R(\theta) ~\overline{s}, ~~
 \overline{e} \rightarrow R(\theta) ~\overline{e},  \]
the angle $\theta$ being arbitrary. This means that all these rotated states of the system are equivalent
and appear entangled with their corresponding environmental partners with equal coefficients. 
 
\vspace*{.4cm}
\noindent
Returning to Zurek's work, his next step is to replace the states $\vert e_{j}>$ by a sum over a 
larger number of states
\begin{equation}
\vert e_{1}> = \frac{1}{\sqrt{n_{1}}}\sum_{l=1}^{n_{1}} \vert e_{l}^{\prime}>, ~~ 
\vert e_{2}> = \frac{1}{\sqrt{n_{2}}}\sum_{l=1}^{n_{2}} 
\vert e_{l}^{\prime \prime}>,
\end{equation}
where $n_{1}$ and $n_{2}$ are integers that could be very large and
\[ <e_{k}^{\prime} \vert e_{j}^{\prime}> = <e_{k}^{\prime \prime} 
\vert e_{j}^{\prime \prime}> = \delta_{jk},~~ <e_{k}^{\prime} \vert e_{j}^{\prime \prime}>=0. \]

\vspace*{.4cm}
\noindent
All in all, Zurek cuts a single environmental state into a number of other states,
keeping the norms properly intact. This procedure looks innocuous, but is it? 

\vspace*{.4cm}
\noindent
The point I wish to make is that in quantum field theory splitting a state into substates
is expected to lead to new physics. Thereby, the modified version {is not} necessarily the same as
the original theory for which the proof was to be given. 

\vspace*{.4cm}
\noindent
As an example consider  
the state of a neutral pion in the quark model. In the original quark model, the neutral pion was
described by the superposition 

\begin{equation}
 \vert \pi^{0} > = \frac{1}{\sqrt{2}}\vert u\overline{u}-d \overline{d} >,
\label{pi}
\end{equation}
where $u$ and $d$ denote the up and down quarks respectively and the bar stands for
the antiparticle. With the discovery that the quarks are ``colored'', i.e., each of
them comes in three equivalent varieties, one had to replace
$u \overline{u}$ by $\frac{1}{\sqrt{3}} \sum_{j=1}^{3} u^{j} \overline{u^{j}}$
and perform a similar replacement for the down quark. Here 
$j$ stands for the color degree of freedom. [This construction is similar to what
Zurek does in his proof.] Thus the neutral pion state becomes
\begin{equation}
  \vert \pi^{0} >= \frac{1}{\sqrt{2}} \frac{1}{\sqrt{3}} \sum_{j=1}^{3} \vert u^{j}~
\overline{u^{j}}-d^{j} ~\overline{d^{j}} >.
\label{pic}
\end{equation}
Note that the $\sqrt{3}$ ensures that the pion state is
properly normalized. Nonetheless, the fact that the ``colorless'' (or one-color, if you prefer) state
in equation (\ref{pi}) has been replaced with three leads to different physics, in spite of the fact 
that color is confined, i.e., it does not leak out. For example, the above modification changes the prediction
of the neutral pion lifetime by a factor of nine!

\vspace*{.4cm}
\noindent
An interesting example, where a state is indeed replaced by a superposition of
other states, is found in the Pauli-Villars regularization scheme \cite{pauli} in quantum electrodynamics. 
In this case the field of the electron is replaced by a sum which includes a number of other
hypothetical fields, with the same quantum numbers as the electron. However, these auxiliary fields 
are taken to be very massive (so that they can not be produced) and appear in the theory with
{\it negative probabilities}. In this manner, one is able to keep the physics intact up to a large
energy scale, beyond which the theory hits the domain of negative probabilities and becomes 
unphysical.

\vspace*{.4cm}
\noindent
The upshot of this note is that splitting a state into substates leads to new physics
in quantum field theory (which encompasses quantum mechanics). The ensuing theory is not, in general,
equivalent to the original one. Therefore, a proof based on
the above cutting procedure is not necessarily conclusive.


\begin{thebibliography}{99}

\bibitem{born1926} M. Born, Z. Physik 37 (1926) 863; 38 (1926) 803

\bibitem{maxS} M. Schlosshauer, ``Decoherence and the Quantum-to-Classical Transition'',
Springer-Verlag, Corrected Third Printing (2008). This book is available on the internet.

\bibitem{zurek2011} W. H. Zurek, Phys. Rev. Lett. 106 (2011) 250402. This article contains
references to Zurek's earlier work.

\bibitem{pauli} W. Pauli and F. Villars, Rev. Mod. Phys. 21 (1949) 434 

\end{thebibliography}
\end{document}